\documentclass[aps,reprint,pra,superscriptaddress]{revtex4-1}
\usepackage[version=3]{mhchem}
\usepackage{graphicx}
\usepackage{floatrow}
\usepackage{amssymb}
\usepackage{amsmath}
\usepackage{commath}
\usepackage{blindtext}
\usepackage{siunitx}
\usepackage{miller}
\usepackage{xcolor}
\usepackage{float}
\usepackage{lipsum}
\usepackage{ulem}

\begin{document}

\title{Spin-orbit semimetal \ce{SrIrO3} in the two-dimensional limit}

\author{D.~J.~Groenendijk}
\email{d.j.groenendijk@tudelft.nl}
\affiliation{Kavli Institute of Nanoscience, Delft University of Technology, P.O. Box 5046, 2600 GA Delft, Netherlands}
\author{C.~Autieri}
\affiliation{Consiglio Nazionale delle Ricerche CNR-SPIN, UOS L'Aquila, Sede Temporanea di Chieti, 66100 Chieti, Italy}
\author{J.~Girovsky}
\author{M.~Carmen Martinez-Velarte}
\author{N.~Manca}
\author{G.~Mattoni}
\author{A.~M.~R.~V.~L.~Monteiro}
\affiliation{Kavli Institute of Nanoscience, Delft University of Technology, P.O. Box 5046, 2600 GA Delft, Netherlands}
\author{N.~Gauquelin}
\author{J.~Verbeeck}
\affiliation{Electron Microscopy for Materials Science (EMAT), University of Antwerp, 2020 Antwerp, Belgium}
\author{A.~F.~Otte}
\affiliation{Kavli Institute of Nanoscience, Delft University of Technology, P.O. Box 5046, 2600 GA Delft, Netherlands}
\author{M.~Gabay}
\affiliation{Laboratoire de Physique des Solides, Bat 510, Universit\'e Paris-Sud, 91405 Orsay, France}
\author{S.~Picozzi}
\affiliation{Consiglio Nazionale delle Ricerche CNR-SPIN, UOS L'Aquila, Sede Temporanea di Chieti, 66100 Chieti, Italy}
\author{A.~D.~Caviglia}
\affiliation{Kavli Institute of Nanoscience, Delft University of Technology, P.O. Box 5046, 2600 GA Delft, Netherlands}
\date{\today}


\begin{abstract}
We investigate the thickness-dependent electronic structure of ultrathin \ce{SrIrO3} and discover a transition from a semimetallic to a correlated insulating state below 4 unit cells. Low-temperature magnetoconductance measurements show that spin fluctuations in the semimetallic state are significantly enhanced while approaching the transition point. The electronic structure is further studied by scanning tunneling spectroscopy, showing that 4 unit cells~\ce{SrIrO3} is on the verge of a gap opening. Our density functional theory calculations reproduce the critical thickness of the transition and show that the opening of a gap in ultrathin \ce{SrIrO3} is accompanied by antiferromagnetic order.
\end{abstract}
\maketitle



Recent advancements in oxide thin film technology have enabled the synthesis of complex materials at the atomic scale. Through interface and strain engineering it is possible to tailor the delicate balance between competing energy scales and control the ground state of quantum materials~\cite{yoshimatsu2010dimensional, zubko2011interface}. 
In the two-dimensional limit, the coordination of constituent ions at the interfaces is reduced, typically yielding a decrease of the electronic bandwidth $W$. At a critical thickness depending on the relative magnitude of $W$ and the Coulomb repulsion $U$, a metal-insulator transition can occur~\cite{hubbard1963electron}. This approach has been applied to study the dimensionality-driven metal-insulator transition (MIT) in 3$d$ transition metal oxides such as \ce{SrVO3} and \ce{LaNiO3}, where a transition from a bulk-like correlated metallic phase to a Mott or static ordered insulating phase occurs in the two-dimensional limit~\cite{yoshimatsu2010dimensional, Boris:2011aa, king2014atomic, scherwitzl2011metal}.

In this Letter, we consider the 5$d$ oxide \ce{SrIrO3} which, in the three-dimensional limit, is a narrow-band semimetal bordering a Mott transition due to a combination of strong spin-orbit coupling (SOC) and electron correlations~\cite{nie2015interplay}. We find that an MIT occurs at a film thickness between 3 and 4 unit cells and study the evolution of the electronic structure across the transition by (magneto)transport and scanning tunneling spectroscopy (STS). The paramagnetic susceptibility is found to be strongly enhanced while approaching the transition point, which is indicative of the opening of a Mott gap and the concomitant enhancement of magnetic order~\cite{imada1998metal}. Our results are supported by first-principles density functional theory (DFT) calculations, which reproduce the critical thickness of the transition and show that the insulating state in the two-dimensional limit is antiferromagnetically ordered. Our study highlights ultrathin \ce{SrIrO3} as a novel platform for engineering the interplay of magnetism and spin-orbit coupling at oxide interfaces.

\ce{SrIrO3} ($n = \infty$) is the only (semi)metallic member of the Ruddlesden-Popper series of strontium iridates \ce{Sr_{$n+1$}Ir_{$n$}O_{$3n+1$}}. On the other end of the series, two-dimensional \ce{Sr2IrO4} ($n=1$) is a Mott insulator with canted antiferromagnetic order. Despite the extended 5$d$ orbitals, narrow, half-filled $J_\mathrm{eff} = 1/2$ bands emerge due to the strong SOC ($\sim0.4\;\mathrm{eV}$) and even a relatively small $U\sim0.5\;\mathrm{eV}$ is sufficient to induce a so-called spin-orbit Mott ground state~\cite{kim2008novel, kim2009phase}. In \ce{SrIrO3}, the effective electronic correlations are smaller due to the three-dimensional corner-sharing octahedral network~\cite{kawasaki2016evolution}, but the strong SOC still causes a significant reduction of the density of states (DOS) at the Fermi level. Together with octahedral rotations that reduce the crystal symmetry, this places the material at the border of a Mott transition and gives rise to an exotic semimetallic state~\cite{nie2015interplay, pallecchi2016thermoelectric}. To study changes in electronic structure between the two end members of the Ruddlesden-Popper series, previous studies have focused on \ce{SrTiO3}/\ce{SrIrO3} superlattices~\cite{kim2014electronic, matsuno2015engineering, kim2016manipulation}. In this system, the crossover from three-dimensional semimetal to two-dimensional insulator was investigated by reducing the number of \ce{SrIrO3} layers. However, it was recently shown that additional hopping channels between the \ce{Ir} atoms are activated by the \ce{SrTiO3} between \ce{SrIrO3} layers, increasing the bandwidth and reducing the effective strength of correlations~\cite{kim2016manipulation}. In the present work, we isolate the effect of dimensionality by studying \ce{SrIrO3} layers of different thickness, providing access to the intrinsic properties of \ce{SrIrO3} in the two-dimensional limit. 



A series of \ce{SrIrO3} films with thicknesses varying from $30$ to $2$ u.c.~were grown by pulsed laser deposition (PLD) on \ce{TiO2}-terminated \ce{SrTiO3}(001) substrates. As described in previous work, we use a \ce{SrTiO3} cap layer to prevent degradation of the film in ambient conditions and enable lithographic processing~\cite{groenendijk2016epitaxial}. Atomic scale characterization of the lattice structure was performed by \ce{Cs}-corrected high angle annular dark field scanning transmission electron microscopy (HAADF-STEM). Hall bars were patterned by e-beam lithography, and the buried \ce{SrIrO3} layer was contacted by \ce{Ar} etching and in-situ deposition of \ce{Pd}/\ce{Au} contacts, resulting in low-resistance Ohmic contacts. Transport measurements were performed in a \ce{He} flow cryostat with a $10\;\mathrm{T}$ superconducting magnet and a base temperature of $1.5\;\mathrm{K}$. Uncapped \ce{SrIrO3} films were transferred in an \ce{N2} atmosphere from the PLD chamber to the low-temperature scanning tunneling microsopy (STM) setup without exposure to ambient conditions. More details regarding the growth and sample characterization can be found in the supplementary material~\cite{[{See supplementary material at http://\ldots for additional information on film characterization, (magneto)transport measurements and first-principles calculations}][{}]suppmat} and in Ref.~\cite{groenendijk2016epitaxial}. First-principles DFT calculations were performed within the Generalized Gradient Approximation using the plane wave VASP~\cite{kresse1999ultrasoft} package and PBEsol for the exchange-correlation functional~\cite{perdew2008restoring} with SOC. The Hubbard $U$ effects on the \ce{Ir} and \ce{Ti} sites were included. To find a unique value of the Coulomb repulsion for the \ce{Ir} 5$d$ states, $U$ was tuned in order to reproduce the experimental semimetallic behaviour at $4\;\mathrm{u.c}$, while we used $J_H = 0.15U$. Using this approach we obtained $U = 1.50\;\mathrm{eV}$, which is in good agreement with the typical values used for weakly correlated \ce{Ir} compounds~\cite{kim2017dimensionality}.

\begin{figure}[ht!]
\includegraphics[width=\linewidth]{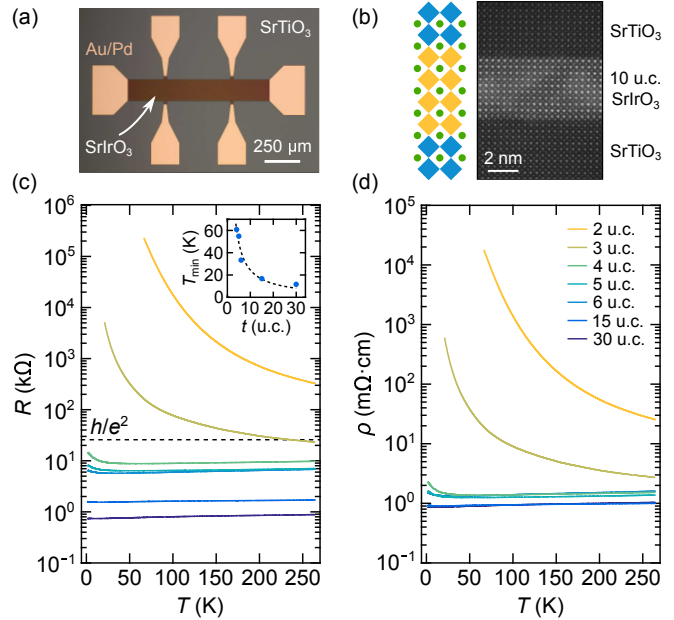}
\caption{\label{Fig1} (a) Optical image of a Hall bar used for (magneto)transport measurements. (b) HAADF-STEM image of a \ce{SrTiO3}/10 u.c.~\ce{SrIrO3}/\ce{SrTiO3} heterostructure. (c) $R(T)$ and (d) $\rho(T)$ curves for films of different thicknesses. The inset shows the temperature of the resistance minimum ($T_\mathrm{min}$) as a function of thickness. The dashed line is a guide to the eye.}
\end{figure}

Figure~\ref{Fig1}(a) shows an optical image of a Hall bar used for transport measurements. The image is taken prior to the removal of the resist mask used to protect the film during the Ar etching step. A HAADF-STEM image of a 10 u.c.~\ce{SrIrO3} film is shown in panel (b), where atomically sharp interfaces with the substrate and the cap layer are visible. The sheet resistance $R$ versus temperature $T$ of \ce{SrIrO3} films with thicknesses $t$ from 30 to 2 unit cells is shown in Fig.~\ref{Fig1}(c). As the film thickness is reduced, $R$ continuously increases and two different regimes can be identified. For $t \geq 4$ u.c., the resistance values are below $25\;\mathrm{k\Omega}$ and the films show metallic behavior. Thinner films ($t \leq 3$ u.c.) have a resistance above $25\;\mathrm{k\Omega}$ and display insulating behavior. Hence, it is apparent that \ce{SrIrO3} films undergo a sharp metal-insulator transition between 4 and 3 u.c., occurring when the sheet resistance crosses $h/e^2 = 25\;\mathrm{k\Omega}$. This is in good agreement with photoemission measurements, which show the disappearance of the Fermi cutoff below 4 u.c.~and the opening of a charge gap~\cite{schutz2017dimensionality}. In two dimensions, the resistance value $h/e^2$ corresponds to the limit $k_\mathrm{F}l_\mathrm{e}\sim1$, where $k_\mathrm{F}$ is the Fermi wavevector and $l_\mathrm{e}$ is the mean free path, marking the transition from weak to strong localization~\cite{licciardello1975constancy}.

In the (semi)metallic regime, the films show bad metallic behavior in the high temperature range, consistent with previous reports~\cite{biswas2014metal, zhang2015tunable, groenendijk2016epitaxial}. The resistance first decreases linearly with temperature until $T_\mathrm{min}$, below which an upturn is observed. In addition, the residual resistance ratio defined as $R(300\;\mathrm{K})/R(T_\mathrm{min})$ is rather low for all thicknesses ($\sim1.2$). Such anomalous metallic behavior is often observed in materials that are bordering a Mott transition. Upon decreasing the film thickness, the temperature of the resistance minimum $T_\mathrm{min}$ increases from $10\;\mathrm{K}$ (30 u.c.) to $60\;\mathrm{K}$ (4 u.c.) [Figure~\ref{Fig1}(c), inset]. By rescaling the curves in panel (c) for the film thickness, we obtain the resistivity $\rho$ as function of temperature as shown in Fig.~\ref{Fig1}(d). In the semimetallic regime, the curves collapse and display similar behavior apart from the increasingly strong upturn at low temperature. Interestingly, the resistance upturn is accompanied by an increase of the Hall coefficient $R_\mathrm{H}$, as shown in the supplementary material~\cite{[{See supplementary material at http://\ldots for additional information on film characterization, (magneto)transport measurements and first-principles calculations}][{}]suppmat}. This is most likely related to the band structure as underscored by angle-resolved photoemission spectroscopy (ARPES) measurements, where multiple heavy hole and light electron bands were identified~\cite{nie2015interplay, liu2016direct}. Since the top energy of several hole bands was measured to lie just below the Fermi level, these bands will be progressively depopulated with decreasing temperature, increasing $R_\mathrm{H}$ and the resistance. 

Transport in ultrathin (2 and 3 u.c.) films occurs in a strongly localized regime with a sheet resistance well in excess of $h/e^2$. For the 3 u.c.~film, the conductivity $\sigma$ can be well described by a variable range hopping (VRH) type of conduction. In this case, electrons hop between localized states and the conductance is given by $\sigma = C\exp[-(T_0/T)^\alpha]$, where $T_0$ depends on the density of localized states and the spread of their wave functions~\cite{brenig1973hopping}. VRH conductivity can be of either Mott or Efros-Shklovskii type, which for a 2D system translates into exponents $\alpha = 1/3$ and $1/2$, respectively~\cite{rosenbaum1991crossover}. The fit to the data yields an exponent $\alpha = 0.57$, which is in good agreement with the latter, suggesting the existence of a Coulomb gap. On the other hand, the $R(T)$ of the 2 u.c.~film can be well fitted by an Arrhenius-type behavior where $R\propto\exp(E_\mathrm{g}/2k_\mathrm{B}T)$, which yields an energy gap of approximately $E_\mathrm{g} = 95\;\mathrm{meV}$.

\begin{figure}[ht!]
\includegraphics[width=\linewidth]{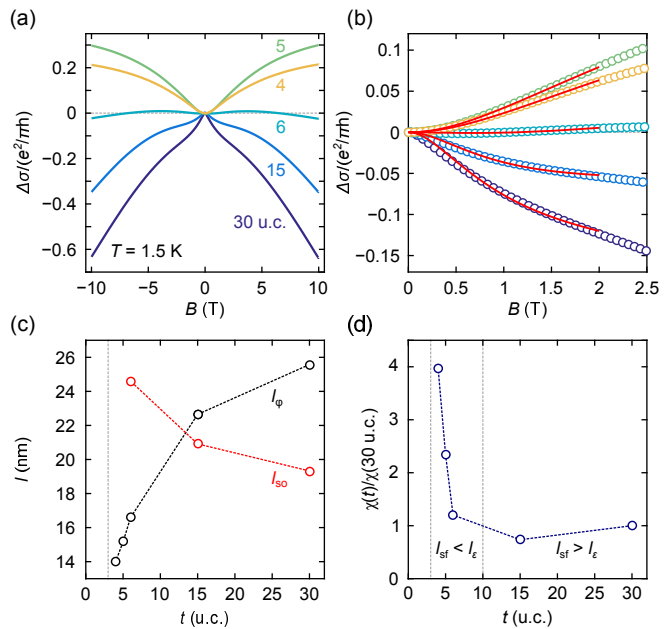}
\caption{\label{Fig2} (a) Magnetoconductance $\Delta\sigma = \sigma(B) - \sigma(0)$ in units of $e^2/\pi h$ measured at 1.5 K in out-of-plane magnetic field for films of different thicknesses. (b) $\Delta\sigma$ fitted by the Maekawa-Fukuyama formula (solid red lines). (c) $l_\mathrm{so}$ and $l_\varphi$ extracted from the fits. (d) Relative susceptibility $\chi(t)/\chi(30\;\mathrm{u.c.})$ versus thickness.}
\end{figure}

To probe changes in the electronic structure and spin relaxation while approaching the transition point, we perform magnetotransport measurements. Figure~\ref{Fig2}(a) shows the out-of-plane magnetoconductance $\Delta\sigma$ in units of $e^2/\pi h$ measured at $1.5\;\mathrm{K}$ for film thicknesses ranging from 30 to 4 unit cells. As shown in the supplementary material, the magnetoconductance is nearly isotropic~\cite{[{See supplementary material at http://\ldots for additional information on film characterization, (magneto)transport measurements and first-principles calculations}][{}]suppmat}. In the limit of large thickness, the magnetoconductance is negative and quadratic and displays a cusp around $B = 0\;\mathrm{T}$ as reported in other works~\cite{biswas2014metal, zhang2015tunable}. However, a crossover from negative to positive values occurs as we approach the MIT. We attribute this behavior to weak (anti)localization, the interference of quantum coherent electronic waves undergoing diffusive motion (in the presence of spin-orbit interaction). In this picture, the magnetic field breaks time-reversal symmetry and destroys the phase coherence of closed paths, suppressing localization effects. To investigate this scenario, we fit the curves with the Maekawa-Fukuyama formula [red lines in Fig.~\ref{Fig2}(b)] in a diffusive regime that describes the change in the conductivity with magnetic field with negligible Zeeman splitting~\cite{hurand2015field}, given by

\begin{align}
\begin{split}
\frac{\Delta\sigma(B)}{\sigma_0} = & -\psi\left(\frac{1}{2} + \frac{B_\mathrm{e}}{B}\right) + \frac{3}{2}\psi\left(\frac{1}{2} + \frac{B_\varphi + B_\mathrm{so}}{B}\right)\\
& - \frac{1}{2}\psi\left(\frac{1}{2} + \frac{B_\varphi}{B}\right) - \ln\left(\frac{B_\varphi + B_\mathrm{so}}{B_\mathrm{e}}\right)\\
& - \frac{1}{2}\ln\left(\frac{B_\varphi + B_\mathrm{so}}{B_\varphi}\right),
\end{split}	
\end{align}

where $\psi$ is the digamma function, $\sigma_0 = e^2/\pi h$ is the quantum of conductance and $B_\mathrm{e}$, $B_\varphi$ and $B_\mathrm{so}$ are the effective fields related to the elastic, inelastic and spin-orbit relaxation lengths, respectively. Since all the films have similar resistivity values, we fix $B_\mathrm{e}$ to $1.2\;\mathrm{T}$, corresponding to an elastic length of approximately $11.7\;\mathrm{nm}$ and a carrier density in the order of $10^{19}\;\mathrm{cm}^{-3}$. This value yields the best fits over the entire thickness range (see supplementary material~\cite{[{See supplementary material at http://\ldots for additional information on film characterization, (magneto)transport measurements and first-principles calculations}][{}]suppmat}) and is consistent with a Drude contribution following our analysis of the semimetallic electronic structure~\cite{manca2017exploring}. For the 30, 15, and 6 u.c.~films, a $B^2$ component was fitted at high fields and subtracted to account for the classical orbital magnetoconductance~\cite{[{See supplementary material at http://\ldots for additional information on film characterization, (magneto)transport measurements and first-principles calculations}][{}]suppmat}. The scattering lengths $l_i$ are related to the effective fields by $B_i = \hbar/4el_i^2$, and their fitted values are shown in Fig.~\ref{Fig2}(c). The lengths are larger than the film thickness, indicating that a 2D model is appropriate. The extracted parameters show a crossover from $l_\varphi>l_\mathrm{so}$ for the thicker samples (30, 15 u.c.) to $l_\varphi<l_\mathrm{so}$ for the thinner ones (6, 5, 4 u.c.), capturing the crossover from negative (weak antilocalization) to positive (weak localization) magnetoconductance as the film thickness is reduced.

A close look at the thickness dependence of $l_\varphi$ reveals deviations from the expected behavior considering only electron-electron corrections to the weak localization expression ($1/l_\varepsilon^2\sim R\log k_\mathrm{F}l_\mathrm{e}$, where $l_\varepsilon$ is the length associated with electron-electron corrections). To correctly describe the physics at play, one needs to include diffusive spin fluctuations which, when sufficiently large, can set the inelastic scattering length, leading to an effective inelastic scattering time given by~\cite{maekawa1981magnetoresistance}

\begin{equation}
\frac{1}{\tau_\varphi} = \frac{1}{\tau_\varepsilon} + \frac{2}{3}\frac{1}{\tau_\mathrm{sf}},
\end{equation}

where $\tau_\varphi$ is related to the energy relaxation time $\tau_{\varepsilon}$ and to the spin fluctuation time $\tau_\mathrm{sf}$ ($l_i^2 = D\tau_i$, where $D$ is the diffusion constant). Since $1/l_\mathrm{sf}^2$ is proportional to the paramagnetic susceptibility $\chi(t)$, we can qualitatively track the variation of $\chi$ by studying the thickness dependence of $l_\varphi$. Figure \ref{Fig2}(d) shows the relative susceptibility $\chi(t)/\chi(30\;\mathrm{u.c.})$ as function of thickness. The increase of $\chi$ at low thicknesses is characteristic of a magnetic transition. We note that the transition from negative to positive magnetoconductance is set by the relative magnitude of $l_\varepsilon$ and $l_\mathrm{sf}$. Near the transition point, $l_\mathrm{sf}<l_\varepsilon$, i.e., spin fluctuations are large, leading to a positive magnetoconductance due to weak localization. In the limit of large thickness, $l_\mathrm{sf}>l_\varepsilon$, $l_\mathrm{so}$. Here, both electron-electron interactions and weak antilocalization contribute to the negative magnetoconductance. 

Structural studies have shown that octahedral coupling at the \ce{SrTiO3}/\ce{SrIrO3} interface suppresses the bulk octahedral rotations in the \ce{SrIrO3} film for $t\leq3$ u.c., enhancing magnetic interactions~\cite{schutz2017dimensionality}. Within this view, the increase of $\chi$ as the film thickness is reduced can be understood as an increased fractional contribution from the less distorted magnetic interfacial region. The film encapsulation could further enhance this effect since it presents two interfaces with the cubic \ce{SrTiO3}.


\begin{figure}[ht!]
\includegraphics[width=\linewidth]{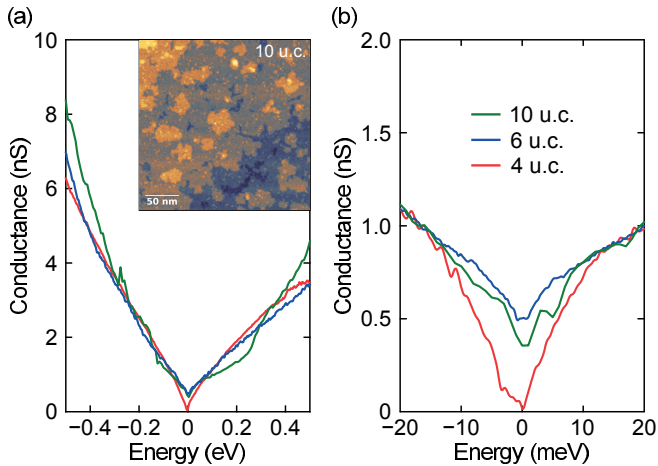}
\caption{\label{Fig3} (a) Differential conductance $(\dif I/\dif V)$ spectra acquired on three different samples with film thicknesses of 4, 6 and 10 unit cells. Inset: STM topographic image of the surface of a 10 u.c.~\ce{SrIrO3} film. (b) $\dif I/\dif V$ spectra measured in a smaller energy range.}
\end{figure}

Further insights on the anomalous behavior in the semimetallic state and the electronic structure near the MIT can be obtained by measuring the DOS across the Fermi energy $E_\mathrm{F}$ by STS measurements. A topographic STM image [inset Fig.~\ref{Fig3}(a)] acquired on a 10 u.c.~\ce{SrIrO3} film shows terraces and steps with height equal to one unit cell, confirming the layer-by-layer growth mode and showing that the surface is single-terminated. Figure \ref{Fig3}(a) shows the differential conductance $(\dif I/\dif V)$ spectra acquired at $4\;\mathrm{K}$ on three different samples with film thicknesses of 4, 6 and 10 unit cells. The spectra taken in the large energy window [Fig.~\ref{Fig3}(a)] show V-shaped behavior with a linear dependence of the DOS for both occupied (negative energies) and unoccupied (positive energies) states. As shown in Fig.~\ref{Fig3}(b), the minimum of the spectra is at zero energy (i.e., at $E_\mathrm{F}$), and while the spectra taken on the 6 and 10 u.c.~films exhibit finite DOS, the 4 u.c.~sample shows zero DOS at $E_\mathrm{F}$. Therefore, the evolution of the DOS at $E_\mathrm{F}$ reflects the approach of the MIT, where the 4 u.c.~film is on the verge of a gap opening.

V-shaped DOS has previously been observed in (1) systems with two-dimensional Dirac surface states such as germanene/\ce{Pt}(111) and graphene/\ce{SiC}~\cite{walhout2016scanning, song2010high} and (2) in the pseudogap phase of lightly-doped Mott insulators such as cuprates~\cite{kohsaka2004imaging, cai2016visualizing}. A Dirac cone is not expected in this system due to the breaking of $n$-glide symmetry by epitaxial constraint, as was shown previously for \ce{SrIrO3} grown on \ce{GdScO3}~\cite{liu2016strain, carter2012semimetal}. However, \ce{Sr2IrO4} exhibits similar V-shaped behavior when doped with \ce{La^{3+}}, showing zero DOS at $E_\mathrm{F}$~\cite{battisti2016universality} as observed for the 4 u.c.~\ce{SrIrO3} film. The resemblance could stem from both \ce{SrIrO3} and doped \ce{Sr2IrO4} being in close proximity to a metal-insulator transition, although on opposite sides of the phase boundary. However, further investigation is required to fully address the exact nature of the V-shaped DOS of \ce{SrIrO3} thin films.


To study the electronic and magnetic structure of \ce{SrIrO3} in the two-dimensional limit and gain additional information about the insulating state, we perform first principles calculations. We first consider how the properties of bulk \ce{SrIrO3} evolve as a function of the Coulomb repulsion $U$. At low $U$, the system shows a nonmagnetic metallic state topologically protected by time-reversal symmetry~\cite{kim2015surface}. Upon increasing the value of $U$, a canted G-type antiferromagnetic (AFM) metallic state with a net in-plane magnetic moment emerges~\cite{matsuno2015engineering}. A further increase of $U$ opens a gap, leading to a G-type AFM insulating state~\cite{zeb2012interplay} like in \ce{SrIrO3}/\ce{SrTiO3} superlattices~\cite{matsuno2015engineering}. Since both $U$ and the breaking of TRS are required to open the gap, the \ce{SrIrO3} thin films can be regarded as insulators located in the intermediate region between a Slater-type and a Mott-type insulator. The same qualitative results were obtained in other \ce{Ir} compounds~\cite{ming2017metal, watanabe2014theoretical}. 

\begin{figure}[ht!]
\includegraphics[width=.8\linewidth]{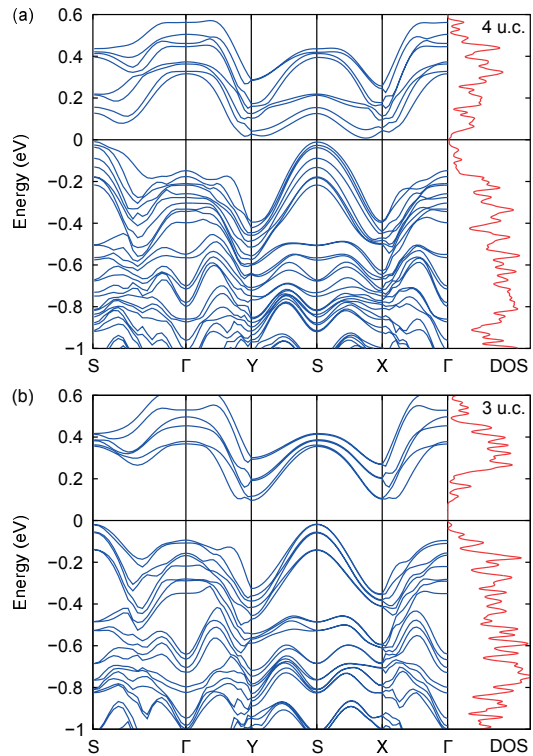}
\caption{\label{fig:BS} Calculated electronic structure for (a) 4 u.c.~and (b) 3 u.c.~\ce{SrIrO3} films on tetragonal \ce{SrTiO3} with $U = 1.50\;\mathrm{eV}$. Right: the corresponding DOS per formula unit as a function of energy.}
\end{figure}

When moving from bulk \ce{SrIrO3} to \ce{SrIrO3}/\ce{SrTiO3} heterostructures, compressive strain, reduction of the bandwidth and an increase of the Coulomb repulsion have to be taken into account. Compressive strain ($\sim1$\%) favors the metallicity~\cite{kim2014electronic} because of the increased bandwidth~\cite{kim2017dimensionality}. The other two effects favor the insulating state~\cite{autieri2016antiferromagnetic} and are both needed to observe the semimetallic or insulating phase in \ce{SrIrO3} ultrathin films. We focused on the thickness range in the vicinity of the MIT and computed the band structure for the 4 and 3 u.c.~films for $U = 1.50\;\mathrm{eV}$, which are shown together with the corresponding DOS in Fig.~\ref{fig:BS}(a) and (b), respectively. 

The reduction of the bandwidth when going from 4 to 3 u.c.~results in a localization of the carriers, and triggers a transition from a semimetallic to an AFM insulating state. Even for a single layer of \ce{SrIrO3} on \ce{SrTiO3} the nonmagnetic case is found to be metallic, and AFM ordering is required for the opening of a gap~\cite{schutz2017dimensionality}. The electronic structure of the 4 u.c.~film shows a gap-closing behaviour, consistent with STS. In the case of 3 u.c.~the gap is 60~meV; its precise value is however crucially dependent on many effects such as octahedral distortions, magnetic order, strain, connectivity and Coulomb repulsion. Near the Fermi level, the DOS is dominated by 5$d$ $t_{2g}$ contribution as in bulk \ce{SrIrO3}. Hence, by reducing the thickness, we approach a state closer to $J_\mathrm{eff}=1/2$ as in \ce{Sr2IrO4}. However, while the $t_{2g}$  unoccupied bandwidth is comparable to \ce{Sr2IrO4}, the occupied part shows a mixed $J_\mathrm{eff} = 1/2$, $3/2$ behavior rather than a pure $J_\mathrm{eff} = 1/2$ picture.

 
In conclusion, we have shown that the spin-orbit semimetal \ce{SrIrO3} can be driven into a correlated insulating state in the two-dimensional limit. At low-temperature, quantum corrections to the conductivity indicate significant changes in scattering mechanisms in the semimetallic regime near the transition point. The divergence of $\chi$ is indicative of the opening of a Mott gap and the concomitant enhancement of magnetic order, in agreement with previous reports of fluctuations in the spin, charge, and orbital degrees of freedom in systems that are approaching a Mott transition~\cite{imada1998metal}. This is corroborated by the near-isotropy of the magnetoconductance, which points towards magnetic scattering in the semimetallic regime. Such isotropy is also observed in thicker films, indicating that there is already a fair amount of magnetic fluctuations in the limit of large thickness, which is understandable in view of the fact that \ce{SrIrO3} is bordering a Mott transition. It is also consistent with previous reports on a diverging magnetic susceptibility at low temperatures and the possibility of exchange enhanced paramagnetism~\cite{pallecchi2016thermoelectric}. The close proximity of \ce{SrIrO3} to a correlated insulating state is further corroborated by STS measurements, showing a V-shaped $\dif I/\dif V$ behavior similar to that of lightly-doped $J_\mathrm{eff} = 1/2$ Mott insulator \ce{Sr2IrO4}. In addition, the 4 u.c.~film reflects the onset of the gap opening as it shows zero DOS at the $E_\mathrm{F}$, being at the border of the MIT. Our DFT calculations reproduce the metal-insulator transition for $U = 1.50\;\mathrm{eV}$ and show that antiferromagnetism develops concomitantly with the opening of a gap.



\begin{acknowledgments}
This work was supported by The Netherlands Organisation for Scientific Research (NWO/OCW) as part of the Frontiers of Nanoscience program (NanoFront), by the Dutch Foundation for Fundamental Research on Matter (FOM). The research leading to these results has received funding from the European Research Council under the European Union's H2020 programme/ERC GrantAgreement n.~[677458]. Support from the French National Research Agency (ANR), project LACUNES No.~ANR-13-BS04-0006-01 is gratefully acknowledged. The authors thank R.~Claessen, P.~Sch\"utz, D.~Di Sante, G.~Sangiovanni and A.~Santander Syro for useful discussions.
\end{acknowledgments}


\bibliographystyle{apsrev4-1} 
\bibliography{References}

\end{document}